\documentclass[12pt]{article}

\usepackage{times}
\usepackage{amssymb}
\usepackage{subfigure}
\usepackage{graphicx}

\makeatletter
\def\blx@maxline{77}
\makeatother
\usepackage[final]{microtype}
\usepackage{csquotes,lmodern}
\usepackage{hyperref}
\usepackage{savesym}

\usepackage[labelfont=bf]{caption}

\renewcommand{\figurename}{Fig.}

\usepackage{filecontents}

\def\aj{\rmfamily{Astron.~J.~}}           
\def\apj{\rmfamily{Astrophys.~J.~}}         
\def\apjl{\rmfamily{Astrophys.~J.~}}        
\def\apjs{\rmfamily{Astrophys.~J~Suppl.~Ser.}}       

\def\aap{\rmfamily{Astron.~Astrophys.}}        

\def\aaps{\rmfamily{Astron.~Astrophys. Suppl.~Ser.}}      
\def\mnras{\rmfamily{Mon.~Not.~R.~Astron.~Soc.~}}     
\def\pasp{\rmfamily{Publ.~Astron.~Soc.~Pacific~}}       
\def\nat{\rmfamily{Nature~}}      

\def\procspie{\rmfamily{Proc.~SPIE~}}

\newcommand{\lya}{Ly$\alpha$}
\newcommand{\MJ}{MACS\,J1423.8+2404}
\newcommand{\HST}{\emph{HST}}

\newcommand{\civ}{\ion{C}{iv}}
\newcommand{\ciii}{\ion{C}{iii}]}
\newcommand{\civlambda}{\ion{C}{iv}$\lambda\lambda1548,1551$}
\newcommand{\ciiilambda}{\ion{C}{iii}]$\lambda\lambda1907,1909$}
\newcommand{\oii}{[\ion{O}{ii}]}
\newcommand{\oiii}{[\ion{O}{iii}]}
\newcommand{\gname}{MACS1423-z7p64}

\newcommand\ion[2]{#1$\;${\scshape{#2}}}

\newcommand\farcs{\mbox{$.\!\!^{\prime\prime}$}}%

\newcommand\othreetwo{$f_{\mathrm{[O\,III]}}/f_{\mathrm{[O\,II]}}$}
 
\newcommand\mubestsixtyeight{$9.6^{+1.8}_{-1.6}$} 
\newcommand\mubestninetyfive{$9.6^{+4.3}_{-2.6}$} 
\newcommand\MUVsixtyeight{$-19.6 \pm 0.2$} 
\newcommand\smasssixtyeight{$3.0^{+1.5}_{-0.8}$} 
\newcommand\SFRsixtyeight{$13.9^{+4.2}_{-3.8}$} 
\newcommand\agesixtyeight{$24^{+16}_{-7}$} 
\newcommand\ssfrsixtyeight{$46.7^{+19.0}_{-17.4}$} 

\newcommand\linefluxunits{$0.41\pm0.06 \, \times 10^{-17} \, \mathrm{erg/s/cm^2}$} 
\newcommand\lineflux{$0.41\pm0.06$} 
\newcommand\lyaEW{$9\pm2$} 
\newcommand\zlya{$z=7.640\pm0.001$}

\newcounter{lastnote}

\author{\normalsize Austin Hoag$^{1\ast}$,  Maru\u{s}a Brada\u{c}$^1$, Michele Trenti$^2$, \\
\normalsize Tommaso Treu$^{3}$, Kasper B. Schmidt$^{4}$, Kuang-Han Huang$^1$, Brian C. Lemaux$^1$, \\
\normalsize Julie He$^1$, Stephanie R. Bernard$^{2}$, Louis E. Abramson$^3$, Charlotte A. Mason$^{3,5}$, \\
\normalsize Takahiro Morishita$^{3,6,7}$,  Laura Pentericci$^8$, and Tim Schrabback$^9$ \\ 
\\
\footnotesize{$^1$Department of Physics, University of California, Davis, CA 95616} \\ 
\footnotesize{$^2$School of Physics, University of Melbourne, VIC 3010, Australia} \\ 
\footnotesize{$^3$Department of Physics and Astronomy, University of California, Los Angeles, CA 90095} \\ 
\footnotesize{$^4$Leibniz-Institut f\"ur Astrophysik Potsdam (AIP), An der Sternwarte 16, 14482 Potsdam, Germany} \\ 
\footnotesize{$^5$Department of Physics, University of California, Santa Barbara, CA 93106-9530, USA} \\ 
\footnotesize{$^6$Astronomical Institute, Tohoku University, Aramaki, Aoba, Sendai 980-8578, Japan} \\ 
\footnotesize{$^7$Institute for International Advanced Research and Education,} \\ 
\footnotesize{Tohoku University, Aramaki, Aoba, Sendai 980-8578, Japan} \\ 
\footnotesize{$^8$INAF Osservatorio Astronomico di Roma, Via Frascati 33,} \\ 
\footnotesize{I-00040 Monteporzio (RM), Italy} \\ 
\footnotesize{$^9$ Argelander-Institut f\"ur Astronomie, Auf dem H\"ugel 71, D-53121 Bonn, Germany} \\
\footnotesize{$^\ast$To whom correspondence should be addressed; E-mail:  athoag@ucdavis.edu.}
}

\title{Spectroscopic Confirmation of an Ultra-Faint Galaxy at the Epoch of Reionization} 

\date{}

\begin{document} 

\maketitle 


\noindent 
\textbf{Within one billion years of the Big Bang, intergalactic hydrogen was ionized by sources emitting ultraviolet and higher energy photons. This was the final phenomenon to globally affect all the baryons (visible matter) in the Universe. It is referred to as cosmic reionization and is an integral component of cosmology. It is broadly expected that intrinsically faint galaxies were the primary ionizing sources due to their abundance in this epoch. However, at the highest redshifts ($z>7.5$; lookback time $13.1$ Gyr), all galaxies with spectroscopic confirmations to date are intrinsically bright and therefore not necessarily representative of the general population. Here we report the unequivocal spectroscopic detection of a low luminosity galaxy at $z>7.5$. We detected the Lyman-alpha ({\lya}) emission line at $\sim10504$\AA{} in two separate observations with MOSFIRE on the Keck I Telescope and independently with the \emph{Hubble} Space Telescope slit-less grism spectrograph, implying a source redshift of $z=7.640\pm0.001$. The galaxy is gravitationally magnified by the massive galaxy cluster MACS\,J1423.8+2404 ($z=0.545$), with an estimated intrinsic luminosity of $\mathrm{M_{AB}}=-19.6 \pm 0.2~\mathbf{mag}$ and a stellar mass of $M^{\star} = 3.0^{+1.5}_{-0.8}\times10^{8}~\mathbf{solar~masses}$. Both are an order of magnitude lower than the four other {\lya} emitters currently known at $z>7.5$, making it probably the most distant representative source of reionization found to date.}

\medskip
\noindent

We originally identified the galaxy (\gname{}, hereafter), as a $z>7$ candidate in \cite{Schmidt+16} using imaging from the Cluster Lensing And Supernova survey with \emph{Hubble} \cite{Postman+12} (CLASH). In \cite{Schmidt+16}, we also reported a tentative $\sim2\sigma$ detection of the hydrogen Lyman-alpha ({\lya}) emission line at $10500\pm50$\AA{} using slit-less spectroscopy from the Grism Lens-Amplified Survey from Space \cite{Schmidt+14,Treu+15a} (GLASS; GO-13459, Principal Investigator PI T.~Treu) . In this work, we present ground-based spectroscopic follow-up of \gname{} obtained with the Multi-Object Spectrometer for Infra-Red Exploration \cite{McLean+12} (MOSFIRE) on the Keck I Telescope on Mauna Kea, HI, on 27 May 2015 (PI M.~Brada\u{c}; 2.3 hrs) and 19 March 2016 (PI M.~Trenti; 1.85 hrs). 

Figure~1 shows the one- and two-dimensional co-added Keck/MOSFIRE spectrum. An emission line is detected at $10504$\AA{} with a signal-to-noise ratio $(S/N)$ of $6.7$. The implied redshift of the galaxy from the MOSFIRE spectrum is {\zlya}, placing it in the top five most distant {\lya} emitters. The wavelength of the line in the MOSFIRE spectrum is consistent with the wavelength of the emission line in the lower resolution \emph{Hubble} Space Telescope (\HST{}) G102 grism spectrum from GLASS (Figure~2). The {\lya} line flux from MOSFIRE is \linefluxunits{}. Using the \HST{} F125W broadband magnitude ($\mathrm{m_{AB}} = 25.32 \pm 0.11$) to estimate the rest-frame ultraviolet (UV) continuum, we measure a rest-frame equivalent width of $W_{\mathrm{Ly}\alpha} = \,$\lyaEW{} \AA{} from the MOSFIRE spectrum. 

We show cutouts of \gname{} in seven \HST{} filters and the \emph{Spitzer}/IRAC $[3.6]\, \mu m$ and $[4.5]\, \mu m$ bands in Figure~3a. The source is detected in F125W, F140W and F160W and exhibits a sharp drop in flux in F105W, consistent with a galaxy spectrum at $z=7.640$ (Figure~3d). We also show a near-infrared false-color \HST{} image of \gname{} and the foreground galaxy cluster lens at $z=0.545$ in Figure~3b. The white line, called the critical line, indicates an infinitesimal band of maximum magnification for sources at $z=7.640$, the redshift of \gname{}. The critical line is very elliptical, a characteristic shared with many other excellent cosmic lenses, and extends to within a few arc seconds of \gname{}. Using the lens modeling software SWunited \cite{Bradac+09}, we determined that the cluster magnifies \gname{} by a factor of {\mubestsixtyeight} ($68\%$ confidence), equivalent to $\sim2.5$ magnitudes. As a result, the intrinsic luminosity of the galaxy at a rest-frame of $1600$\AA{} is $L_{UV}=0.4^{+0.2}_{-0.1}L^{\star}_{UV}$, i.e. $40\%$ of the characteristic luminosity at $z\sim8$, where we adopted $M^{\star}_{UV} = -20.63\pm0.36$ from \cite{Bouwens+15} to determine $L^{\star}_{UV}$. We list all photometric and spectroscopic properties of \gname{} in Table 1. 

Using the longer wavelength G141 GLASS spectra, we can rule out the {\oii} interpretation of the emission line at $z=1.818$; if the emission line at $10504$\AA{} was part of the {\oii} doublet, then the {\oiii} pair would have very likely been detected in the G141 spectra (Figure~2; see also Supplementary Information). The {\oiii} pair is not detected in either P.A., however. Though a sky line in the MOSFIRE spectrum prevents us from reliably measuring the asymmetry often associated with {\lya} emission or definitely ruling out the presence of a doublet, photometric constraints also strongly favor the {\lya} interpretation at {\zlya}. 

Numerous searches for {\lya} at $z>7$ have been carried out, revealing an extremely low {\lya} emitting fraction \cite{Schenker+12,Treu+13,Pentericci+14,Schmidt+16}. We targeted \gname{} along with 9 other $z\gtrsim7$ candidates in \MJ{} using the selection technique discussed by \cite{Schmidt+16} and as part of a larger ongoing campaign to follow up {\lya} emitter candidates from the GLASS program. Of the 9 candidates targeted in \MJ{}, only 1 other target with a marginal {\lya} detection in the GLASS spectra was observed to the same depth as \gname{}, with no evidence of a detection. Therefore, that candidate is likely not a real {\lya} emitter, highlighting the importance of follow-up. A full discussion of the ground-based follow up of the GLASS {\lya} candidates will be presented by Hoag et al. (in preparation).

{\lya} emission has been detected with $S/N>5$ at $z>7.5$ in only four other sources, all of which are $L_{UV} > L^{\star}_{UV}$ sources \cite{Fink+13,Oesch+15,Zitrin+15b,Song+16}, naturally suggesting a relationship between rest-frame $UV$ luminosity and {\lya} visibility. It has been proposed by \cite{Zitrin+15b, Castellano+16} that the preferential success for bright sources could be due to brighter, more massive galaxies tracing the centers of the largest ionized regions, allowing {\lya} to redshift out of resonance before encountering intergalactic hydrogen. This model is often referred to as ``inside-out'' reionization, and is favored by most recent simulations \cite{Bauer+15,Hutter+16}. The detection of {\lya} from a single faint galaxy at $z>7.5$ is not necessarily at odds with this scenario. The size of ionized regions at $z\sim8$ is typically much larger than the \HST{} WFC3/IR field of view (FOV) \cite{Wyithe+2004} such that more massive galaxies part of the same ionized region could have escaped our detection. Within the GLASS FOV, several other {\lya} candidates also exist, though without full depth follow-up \cite{Schmidt+16}.

Another mechanism may be contributing to the visibility of {\lya} from \gname{}. At least 3 out of the 4 other {\lya} emitters at $z>7.5$ exhibit extremely red rest-frame optical (observed mid-infrared) color, i.e. $([3.6]\, \mu m-[4.5]\, \mu m) > 0.5$. For the 4th {\lya} emitter presented by \cite{Song+16}, the mid-infrared color was not measured due to contamination. Bright {\oiii}$\lambda\lambda4959,5007$ and H$\beta$ emission lines--which fall in the \emph{Spitzer}/IRAC $[4.5]\, \mu m$ band at $z\gtrsim7.1$ and are associated with vigorous star formation---may be responsible for such red colors \cite{Labbe+13,RobertsBorsani+16}. The strength of {\oiii} emission has been linked to hard ionization fields in $z=3-4$ galaxies by \cite{Nakajima+16} and tentatively at higher redshift by \cite{Stark+16}, which could also increase the {\lya} visiblity by clearing away the neutral hydrogen in the local environment. \gname{} is only detected with confidence in the $[4.5]\, \mu m$ band, such that we measure the mid-infrared color as a $3\sigma$ lower limit: $([3.6]-[4.5]) > 0.19$. Therefore, it is possible that the galaxy may exhibit a similarly red color to the other systems mentioned above. If this were the case, it could indicate a hard ionizing spectrum, explaining why we are able to detect {\lya} from this source despite its intrinsically low luminosity, $L_{UV}=0.4^{+0.2}_{-0.1}L^{\star}_{UV}$, and stellar mass, {\smasssixtyeight}$\times10^8~\mathrm{solar~masses}~(M_{\odot})$. \gname{} also harbors a very young stellar population ({\agesixtyeight}$~\mathrm{Myr}$), a property it shares with the red mid-infrared galaxies at $z>7.5$.  

An indication that \gname{} may be vigorously forming stars is its large specific star-formation rate (sSFR) implied from spectral energy distribution (SED) fitting: {\ssfrsixtyeight}$\mathrm{Gyr}^{-1}$. This is comparable to the sSFRs of the 3 {\lya} emitters at $z>7.5$ with measured strong rest-frame optical colors. \cite{Stark+16} noted that the red $[3.6]-[4.5]\, \mu m$ color at $z>7.1$ may be due in part to high sSFR. Therefore, the large sSFR implied from SED fitting could imply a harder ionizing spectrum, allowing {\lya} to escape.  

The \textit{James Webb Space Telescope} (JWST) will be sensitive to the rest-frame UV and optical spectrum of \gname{}, making this galaxy an excellent candidate for future follow-up. With modest exposure times, JWST spectroscopic observations could constrain the hardness of the spectrum through rest-frame UV line strengths and measure the {\oiii}/H$\beta$ strengths to determine the nature of the flux excess in the \emph{Spitzer}/IRAC $[4.5]\, \mu m$ band.

\clearpage

\begin{table*}
\vspace{-1in}
        \begin{center}
	\begin{tabular}{||ccc||} 
        \hline
        \multicolumn{3}{|c|}{Photometry } \\
        \hline\hline
$\alpha_{\mathrm{J2000}}$ & ($^{\circ}$) & 215.942406 \\
$\delta_{\mathrm{J2000}}$ & ($^{\circ}$) & 24.069655 \\
$\mu_{\mathrm{best}}$ & & {\mubestsixtyeight} \\        
$M_{1600} - 2.5 \mathrm{log}_{10}(\mu / \mu_{\mathrm{best}})$$^{a}$ & (mag) & {\MUVsixtyeight} \\
$M_{\star} \times \mu / \mu_{\mathrm{best}} $ & ($10^{8} M_{\odot}$) & {\smasssixtyeight} \\
SFR$ \times \mu / \mu_{\mathrm{best}} $ & $M_{\odot} \, \mathrm{yr}^{-1}$ & {\SFRsixtyeight} \\
Age & (Myr) & {\agesixtyeight} \\
\hline
F435W &     (mag) &   $>26.70$ \\
F475W &    (mag) &    $>26.77$ \\
F555W &    (mag) &    $>27.19$ \\ 
F606W &    (mag) &    $>27.16$ \\ 
F775W &    (mag) &    $>26.48$ \\ 
F814W &    (mag) &    $>27.50$ \\ 
F850LP &   (mag) &    	$>26.24$ \\ 
F105W &    (mag) &    $26.46 \pm 0.24$ \\ 
F110W &    (mag) &    $25.79 \pm 0.11$ \\ 
F125W &    (mag) &    $25.32 \pm 0.11$ \\
F140W &    (mag) &    $24.99 \pm 0.06$ \\
F160W &    (mag) &    $25.03 \pm 0.10$ \\
$[3.6] \, \mu m$ &      (mag) &     $>24.69$ \\
$[4.5] \, \mu m$ &      (mag) &     $24.50\pm0.27$ \\\hline
\multicolumn{3}{|c|}{Keck/MOSFIRE Spectroscopy } \\
\hline
$z_{\mathrm{Ly}\alpha}$ & & $7.640 \pm 0.001$ \\
$t_{\mathrm{exp}}$  &  (hr) & 4.15  \\
$f_{\mathrm{Ly}\alpha}^{\mathrm{MOSFIRE}}$ & ($10^{-17}$ erg\,s$^{-1}$\,cm$^{-2}$) & \lineflux{} \\
$W_{\mathrm{Ly}\alpha}^{\mathrm{MOSFIRE}}$ & (\AA) &  \lyaEW{}  \\
\hline
\multicolumn{3}{|c|}{Grism Spectroscopy } \\
\hline
$f_{\mathrm{Ly}\alpha}^{\mathrm{PA8}}$ & ($10^{-17}$ erg\,s$^{-1}$\,cm$^{-2}$) & $<1.35$ \\

$f_{\mathrm{Ly}\alpha}^{\mathrm{PA88}}$ & ($10^{-17}$ erg\,s$^{-1}$\,cm$^{-2}$) & $1.20\pm0.50$ \\

$f_{\mathrm{CIV}}^{\mathrm{PA8}}$ & ($10^{-17}$ erg\,s$^{-1}$\,cm$^{-2}$) & $<1.17$ \\

$f_{\mathrm{CIV}}^{\mathrm{PA88}}$ & ($10^{-17}$ erg\,s$^{-1}$\,cm$^{-2}$) & $<1.23$  \\

$f_{\mathrm{CIII}}^{\mathrm{PA8}}$ & ($10^{-17}$ erg\,s$^{-1}$\,cm$^{-2}$) & $<1.08$ \\

$f_{\mathrm{CIII}}^{\mathrm{PA88}}$ & ($10^{-17}$ erg\,s$^{-1}$\,cm$^{-2}$) & $<1.23$  \\

$W_{\mathrm{Ly}\alpha}^{\mathrm{PA8}}$ & (\AA) & $<30$ \\

$W_{\mathrm{Ly}\alpha}^{\mathrm{PA88}}$ & (\AA) & $27 \pm 11$  \\

$W_{\mathrm{CIV}}^{\mathrm{PA8}}$ & (\AA) & $<30$ \\

$W_{\mathrm{CIV}}^{\mathrm{PA88}}$ & (\AA) & $<33$ \\

$W_{\mathrm{CIII}}^{\mathrm{PA8}}$ & (\AA) & $<21$ \\

$W_{\mathrm{CIII}}^{\mathrm{PA88}}$ & (\AA) & $<24$  \\

\hline
\end{tabular}
    \end{center}
    \caption{
      Photometric and spectroscopic properties of \gname{}. \HST{} magnitudes are normalized by the SExtractor MAG AUTO value in F160W, but colors are measured with MAG ISO. All uncertainties are $1\sigma$, whereas upper limits are $3\sigma$. Equivalent widths are calculated using the measured \HST{} magnitudes to estimate the continuum flux density. F125W, F160W and F140W are used in the EW calculations for {\lya}, {\civ} and {\ciii}, respectively.$^a \,M_{1600}$ is the rest-frame absolute UV magnitude at $1600$\AA{}, calculated from the F140W magnitude.}
\end{table*}
  
\clearpage

\begin{figure}[h]
\centering
\includegraphics[width=\textwidth]{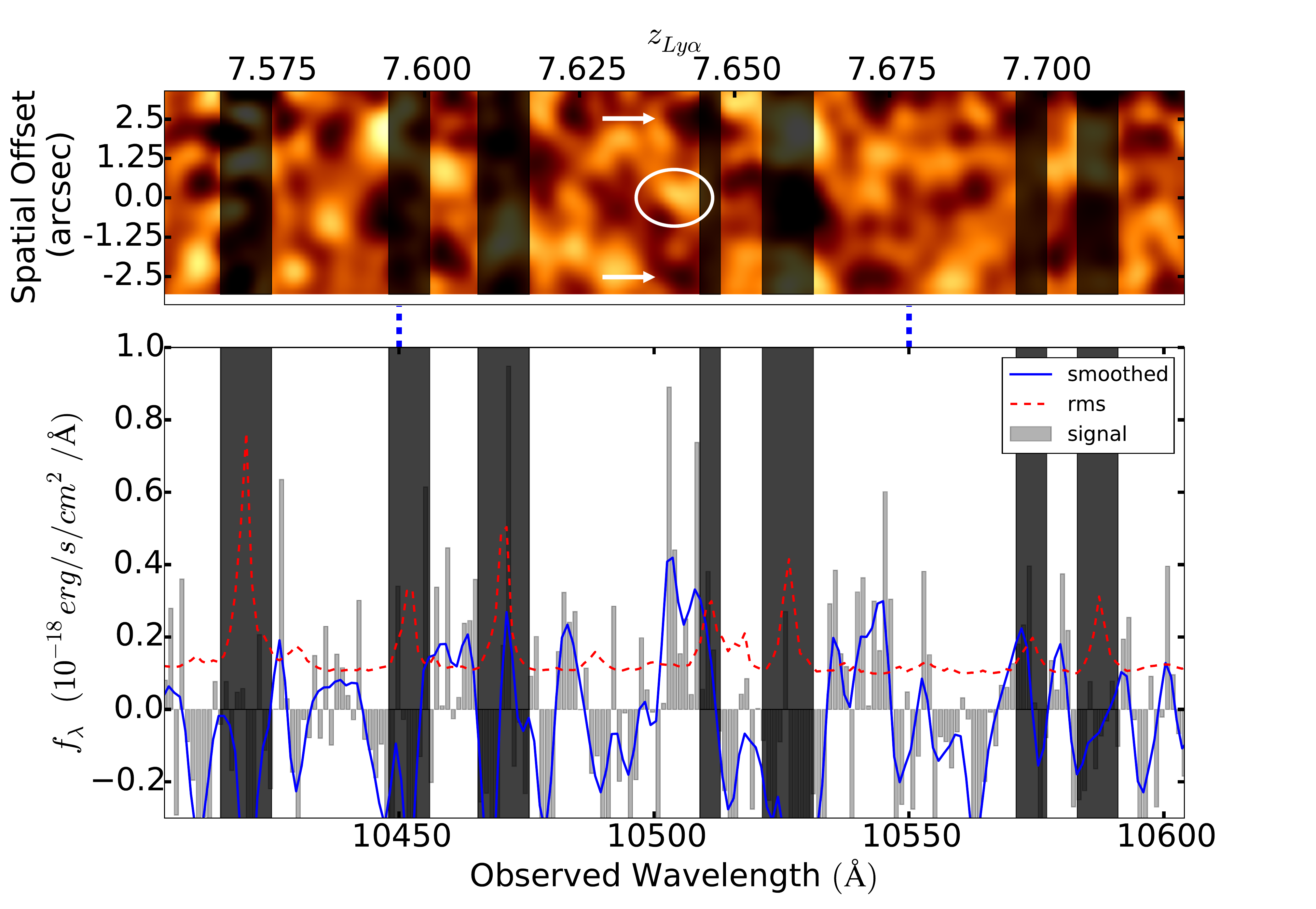}
\caption{Integrated $S/N= 6.7$ detection of the {\lya} emission line at $z=7.640$ with Keck/MOSFIRE. \textbf{Top:} Two-dimensional full-depth (4.15 hr) co-added spectrum. The spectrum has been smoothed to the atmospheric seeing of the observations. The emission line is shown inside the white circle for reference, and the white arrows mark the locations of both negative residuals, which appear at the expected locations and with the expected intensities from the dither pattern. \textbf{Bottom:} The one-dimensional spectrum extracted from the above two-dimensional spectrum. The signal and rms noise are shown with shaded grey and dotted red lines, respectively. Both are unsmoothed and extracted using the same seeing-matched aperture. The rms noise was obtained from the stacked two-dimensional rms spectrum of the two independent MOSFIRE data sets. The blue line is the signal once smoothed to the MOSFIRE spectral resolution. Atmospheric emission lines are masked out in both panels with dark gray vertical bands. The two dashed vertical blue lines between the top and bottom panels represent the $68\%$ confidence interval for the {\lya} wavelength from the GLASS spectra.}
\end{figure} 

\begin{figure}
\centering
\includegraphics[width=\textwidth]{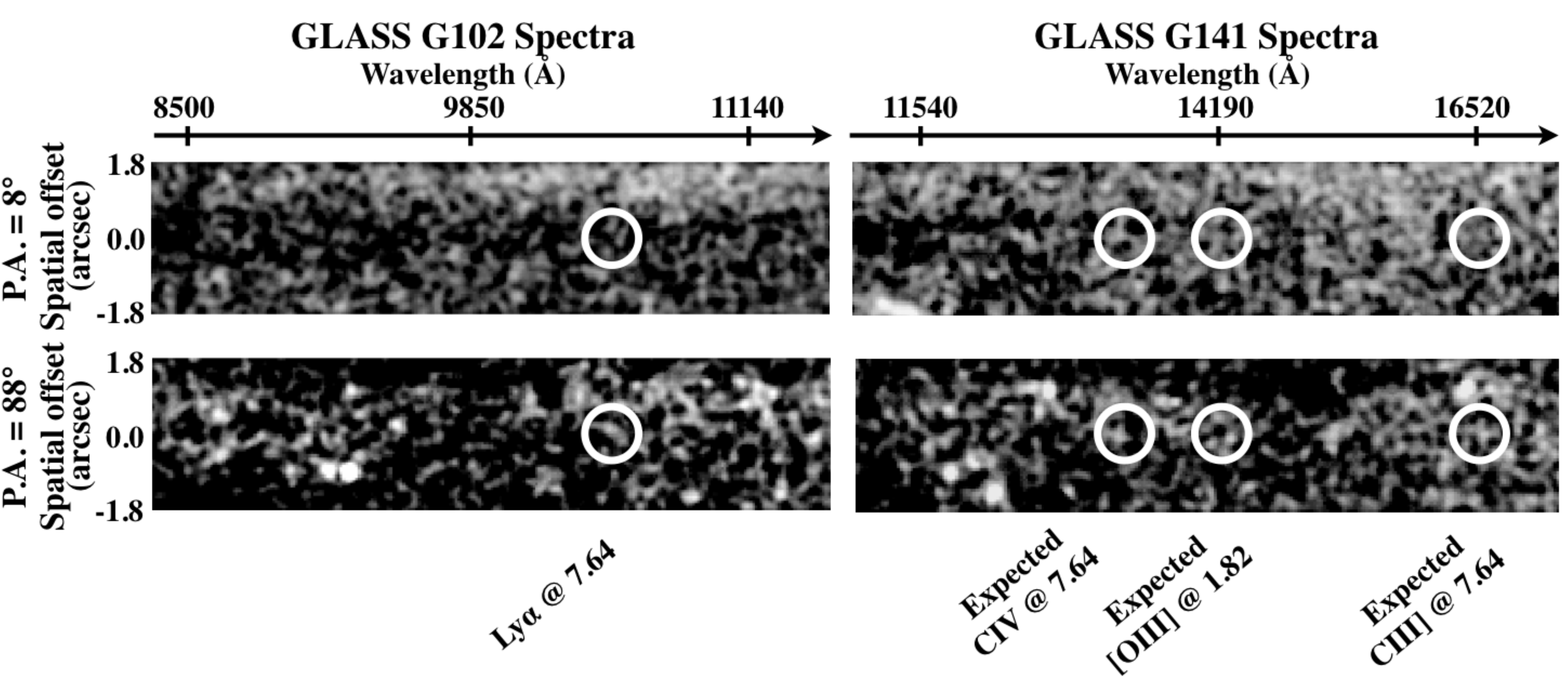}
\caption{\HST{} G102 (left) and G141 (right) two-dimensional contamination-subtracted spectra from the GLASS program \cite{Schmidt+14,Treu+15a}. The two different P.A.s ($8^{\circ}$ and $88^{\circ}$) are shown in the top and bottom panels, respectively. {\lya} emission is detected with $S/N = 2.4$ at $10500\pm50$\AA{} in the P.A.=$88^{\circ}$ G102 spectrum, and an upper limit is obtained in the  P.A.=$8^{\circ}$ spectrum, likely due to contamination. The white circles denote the observed {\lya} and the predicted {\civlambda} and {\ciiilambda} emission lines at $z=7.640$. Marginal flux excesses ($<3\sigma$) are observed for {\civ} and {\ciii} in the P.A.=$88^{\circ}$ G141 spectrum, but follow-up spectroscopy is needed to confirm or deny these features. Also shown is the expected location of the {\oiii} pair at $z=1.818$, the redshift if the {\lya} line was instead {\oii}. {\oiii} is not detected assuming $z=1.818$, which provides strong evidence against the {\oii} interpretation of the line because typically {\othreetwo} $\geq1.5$ for low-mass galaxies at this redshift \cite{Henry+13}. }
\end{figure}

\begin{figure}
\centering
\includegraphics[width=\textwidth]{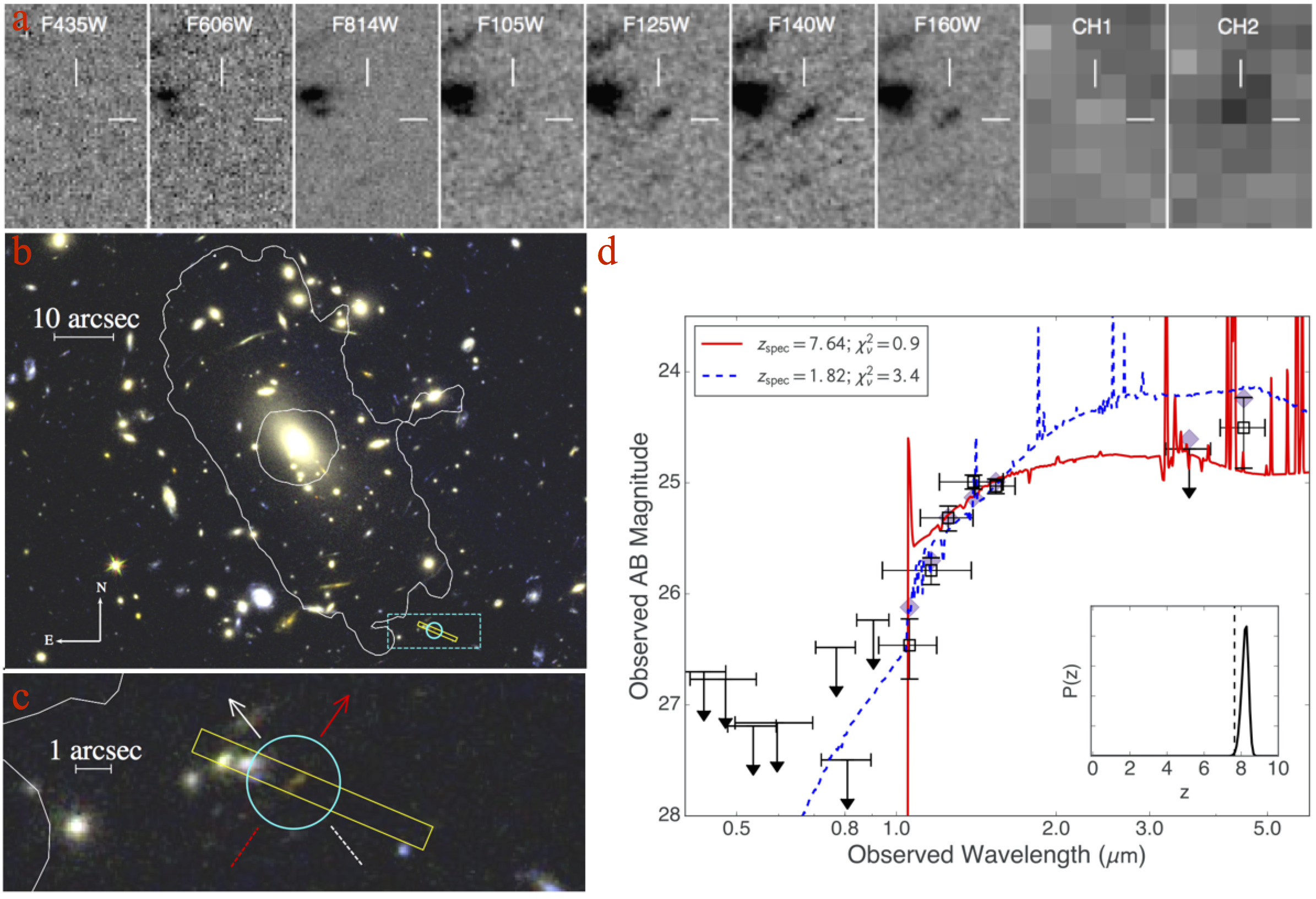}
\caption{(a) $3''$ x $5''$ \HST{} and background-subtracted \emph{Spitzer}/IRAC (CH1 and CH2) postage stamps of \gname{}. Black represents positive signal. The source is detected in F125W, F140W and F160W and not detected in any optical bands. The marginal detection in F105W is consistent with a rapid drop in flux around $1.05\, \mu m$, just blue-ward of {\lya}. The source is detected in IRAC CH2 but not in CH1. (b) \HST{} near-IR false-color image of the galaxy cluster {\MJ} ($z=0.545$), showing the location of \gname{} (cyan circle) relative to the critical line (white line) and the MOSFIRE slit (yellow rectangle). (c) Close-up of region inside dotted cyan rectangle from (b). The dispersion directions from the two GLASS P.A.s are shown by the red (P.A.=$8^{\circ}$) and white (P.A.=$88^{\circ}$) arrows. (d) Observed broadband flux densities (squares) and $3\sigma$ upper limits (downward arrows) from $\sim 0.4-5~\mu m$. Vertical error bars show the $1\sigma$ flux density errors and horizontal error bars show the effective width of each filter. We also show the best-fit 
galaxy spectral energy distributions (SEDs) when redshifts are fixed at the Ly$\alpha$ redshift $z=7.640$ 
(red solid line) and at the hypothetical {\oii} redshift $z=1.818$ (blue dashed line). The
flux densities predicted by the best-fit $z=7.640$ galaxy SED are shown as purple diamonds. The photometric redshift probability
density function obtained by allowing the galaxy redshift to vary is shown in the inset in the lower-right corner. The vertical dashed line in the inset marks the
Ly$\alpha$ redshift, $z=7.640$. }
\end{figure}

\clearpage
  
\section*{Methods}
\subsection*{Imaging Data, Photometry, and Stellar Population Modeling} 

We use the multi-band \emph{HST} imaging data for \MJ{} from CLASH
to constrain the spectral energy distribution (SED) of the galaxy within $\sim0.4-1.7~\mu m$. CLASH mosaics in $0\farcs03$/pixel are resampled onto a 
$0\farcs06$/pixel grid before automatic source detection in F160W, and multi-band 
photometry is performed using SExtractor \cite{Bertin+96} in dual-image mode. Magnitudes in each \HST{} band are measured from the SExtractor ISO magnitude, normalized to the SExtractor AUTO magnitude in F160W. To
mitigate contamination from the extended wings of brightest cluster galaxies 
(BCGs) and intra-cluster light (ICL), we model the BCGs and ICL with smooth light profiles
and subtract them from the images following the steps in \cite{Merlin+16}.
Our simulations with artificial sources suggest that for a point source of 
roughly 25th magnitude, this foreground-subtraction procedure 
reduces the total flux measurement biases by $\sim 0.04$ mag in WFC3/IR filters. 

Colors between \emph{HST} F160W and \emph{Spitzer}/IRAC $3.6~\mu m$ and 
$4.5~\mu m$ bands are measured from the deep \emph{Spitzer}/IRAC images taken
by the \emph{Spitzer} UltRa Faint SUrvey Program \cite{Bradac+14} (SURFSUP).
We measure the galaxy's \emph{Spitzer} flux densities
using \textsc{T-PHOT} \cite{Merlin+15}, which
employs a template-fitting approach to measure accurate colors
between images with different angular resolutions. Empirical 
\emph{Spitzer}/IRAC point-spread functions 
(PSFs) are derived from the stacked stellar images within the \emph{Spitzer}/IRAC coverage; these
PSFs have full widths at half maximum consistent with those published by the IRAC handbook \cite{Bradac+14}.

From the broadband flux densities measured using the above procedure, we
derive constraints on the stellar mass, star formation rate, and dust content
of the galaxy by fitting stellar population synthesis model SEDs. We adopt
the model SEDs from \cite{BC03} with the stellar initial mass function from \cite{Chabrier03} and a constant star formation history. We explored three values of metallicity, $Z = 0.02\,Z_\odot$, $Z= 0.2\,Z_\odot$, and $Z = Z_\odot$, finding that $Z = 0.02\,Z_\odot$ produced the best fit to the data. 
Nebular emission lines and continuum 
are added to the models based on their Lyman continuum flux as in 
\cite{Huang+16a}. We explored two dust attenuation curves, the ``SMC'' \cite{Pei92} and ``Calzetti'' \cite{Calzetti+00}, finding that the SMC dust attenuation curve provided a better fit to the data. In both models, the stellar and nebular emission are parameterized as in \cite{Calzetti+00} by $E(B-V)_s$ and $E(B-V)_{\rm{neb}}$, respectively, related by $E(B-V)_s=0.44E(B-V)_{\rm{neb}}$, and derived from a comparison between Balmer line ratios and the reddening of stellar continua. Larger dust attenuation of the nebular emission relative to the stellar continuum can also be implemented in this way for the SMC dust curve since the SMC and Calzetti dust attenuation curves have very similar shapes at rest-frame optical wavelengths.  

The best-fit model SED at the {\lya} 
redshift $z=7.640$ is shown as the red solid line in Figure~3d,
while the bottom-right inset shows the photometric redshift probability 
distribution when the redshift of the model SED is allowed to vary. We also show
the best-fit model SED at $z=1.818$ (the hypothetical {\oii} redshift)
as the blue dashed line, and we find that the $z=7.640$ solution is strongly favored 
over the $z=1.818$ solution. 

From the best-fit
model SED at $z=7.640$ and a magnification $\mu=\,$\mubestsixtyeight, 
we infer the galaxy's \emph{intrinsic} stellar mass and star formation rate to be
\smasssixtyeight$\,\times 10^8~M_\odot$ and 
\SFRsixtyeight$~M_\odot$/yr, respectively, resulting in a sSFR of 
\ssfrsixtyeight $\,\mathrm{Gyr}^{-1}$.  
Error bars represent 68\% confidence intervals
that include magnification uncertainty (except for sSFR, which is independent of $\mu$);
they are derived from Monte Carlo resampling of the photometry and 
refitting using the same library of model SEDs \cite{Huang+16a}.

The inferred age of the stellar population is young (\agesixtyeight $\,\mathrm{Myr}$), despite a moderate amount of dust extinction at rest-frame 1600 \AA\ ($A_{1600}\approx1.8$ mag when
the SMC curve is used). The combination of very young age and moderate amount of dust extinction has been inferred for other high-$z$ LAEs \cite{Fink+13}. Given the inherent degeneracies between age, dust, and metallicity, it is difficult to further constrain these parameters without independent measurements. We experimented with stellar population templates including instantaneous bursts of star formation at various ages (100, 200, 500 and 700 $\, \mathrm{Myr}$) in combination with the constant star-formation history templates used in our fiducial best-fit SED. The addition of instantaneous burst templates did not result in a better fit to the data than the constant star-formation history templates alone. Therefore, we do not find strong evidence to suggest a hidden, old stellar population for this galaxy.

\subsection*{Gravitational Lens Model of the Galaxy Cluster} 

The model presented here is a revision of the grid-based lens model of \MJ{} first appearing in \cite{Vulcani+15}. We use the available strong lensing constraints from the previous models as well as improved weak lensing constraints from the \HST{} Advanced Camera for Surveys (ACS) F814W filter. Briefly, the lens modeling technique \cite{Bradac+05,Bradac+09} reconstructs the gravitational potential on a non-uniform grid via a $\chi^2$ minimization of strong and weak lensing terms. The method converges much more readily when provided with a reasonable initial model. Our initial model consists of two non-singular isothermal ellipsoids (NIEs) at the locations of the two BCGs and non-singular isothermal spheres (NIS) at the locations of the 10 brightest cluster members within the HST field of view. The velocity dispersions of the two NIEs and NIS cluster members are determined by attempting to broadly reproduce the strong lensing positions. The NIS cluster members are assigned weights based on their F105W magnitudes, whereas the NIE velocity dispersions are chosen individually. We also include in the model a faint ($M_{F105W} = 23.4$ mag) galaxy because it is at a small projected separation ($1\farcs5$) from \gname{}. The galaxy likely belongs to the cluster based on its photometric redshift ($z_{\mathrm{phot}} = 0.46\pm0.08$). Including this galaxy in the initial model does not significantly affect the fit to the data, but it increases the magnification of \gname{} by $\sim10\%$. Thus, we include it in the initial model since its effect on \gname{} is significant. 

The lens model is constrained by three multiple image systems with spectroscopic redshifts reported by \cite{Limousin+10} and one with a photometric redshift identified by \cite{Zitrin+15a}. One of the spectroscopic systems is at $z=1.779$ and the other two are both at $z=2.84$. The errors on the spectroscopic redshifts were not provided by \cite{Limousin+10}. The two systems at $z=2.84$ may be two different knots of the same source, but we include them as two separate systems to increase the number of constraints on the model. We confirm the redshift of the system at $z=1.779$ with the detection of an emission line at $10370\pm50$\AA{} in the GLASS spectroscopy of all three images in the system, consistent with \oii{} at $z=1.78\pm0.01$. We find no other significant spectral features in the GLASS spectroscopy of these three images or of the other potential images in other systems. The photometric redshift for the final system is $z_{\mathrm{phot}} = 1.97\pm0.15$, which we fix to $z=1.97$ in the model. To generate the weak lensing catalog we use the pipeline described by \cite{Schrabback+10,Schrabback+16}. The source density of weak lensing galaxies used in the model is $\sim120~\mathrm{arcmin^{-2}}$, and the mean ratio of angular diameter distances ($\beta_{s}$) after making a photo-$z$ cut at $z>z_{\mathrm{cluster}}+0.1$ is $\langle \beta_{s} \rangle = 0.55\pm0.01 \,(68\% \, \mathrm{confidence}$). 

The magnification of \gname{} from our model is {\mubestninetyfive} $[95\% \,\mathrm{confidence}]$. Two previous lens models using the same strong lensing constraints are publicly available, which are the Zitrin LTM-Gauss and NFW v2 models \cite{Zitrin+15a}, and predict the magnification of \gname{} to be $17.2^{+2.6}_{-2.5}$ and $10.6^{+1.1}_{-2.0} \,[95\% \,\mathrm{confidence}]$, respectively. At the 95\% confidence level, our model agrees with the NFW v2 model, but disagrees with the LTM-Gauss model. All three models agree that \gname{} is magnified by $\mu \gtrsim 10$. The critical line from our lens model at $z=7.640$ is shown in Figure~3b. As in our model, \gname{} is outside of the critical line at $z=7.640$ in the two \cite{Zitrin+15a} models, and likely not multiply imaged. However, we search for additional images in the case that \gname{} has a counter-image that is located inside the $z=7.640$ critical line. No additional images are predicted by any of the models. Throughout this work we adopt the magnification and uncertainties from our model only.

\section*{Acknowledgements}
AH and this work were supported by NASA Headquarters under the NASA Earth and Space Science Fellowship Program - Grant ASTRO14F-0007. Data presented herein were obtained at the W.M. Keck Observatory, which is operated as a scientific partnership among the California Institute of Technology, the University of California and the National Aeronautics and Space Administration. The Observatory was made possible by the generous financial support of the W.M. Keck Foundation. The authors thank Luca Rizzi and Marc Kassis for help with MOSFIRE observations and data reduction. The authors wish to recognize and acknowledge the very significant cultural role and reverence that the summit of Maunakea has always had within the indigenous Hawaiian community.  We are most fortunate to have the opportunity to conduct observations from this mountain. This work is also based on observations made with the NASA/ESA Hubble Space Telescope, obtained at the Space Telescope Science Institute (STScI), which is operated by the Association of Universities for
Research in Astronomy, Inc., under NASA contract NAS5-26555 and NNX08AD79G and ESO-VLT telescopes. Support for GLASS ({\HST}-G0-13459) was provided by NASA through a grant from STScI. We are very grateful to the staff of the Space Telescope for their assistance in planning, scheduling and executing the observations, and in setting up the GLASS public release website. Support for this work was also provided by NASA through an award issued by JPL/Caltech and through {\HST}-AR-13235, {\HST}-GO-13177, {\HST}-GO-10200, {\HST}-GO-10863, and {\HST}-GO-11099 from STScI. Observations were also carried out using the Spitzer Space Telescope, which is operated by the Jet Propulsion Laboratory, California Institute of Technology under a contract with NASA. Support for this work was also provided by NASA through a Spitzer award issued by JPL/Caltech. 

\clearpage

\newpage

\begin{center}
\section*{\Large Supplementary Information}
\end{center}

\renewcommand{\figurename}{Supplementary Figure}
\setcounter{figure}{0}  

\subsection*{Reduction and calibration of the Keck/MOSFIRE spectroscopic data} 

We targeted \gname{} in the Y-band with two distinct multi-object slit-masks, which were created using the MOSFIRE Automatic GUI-based Mask Application (MAGMA). The slit masks used on 27 May, 2015 (PI Brada\v{c}; 2.30 hrs) and 19 March 2016 (PI Trenti; 1.85 hrs) had the same position angle, but some of the filler slits differed between the masks. We noticed an astrometric shift of $\sim0\farcs1$ between the two mask files created by MAGMA when overlaying them on an \HST{} image. This offset is smaller than one MOSFIRE pixel, so we do not expect that it influences our results obtained from stacking the spectra. On 27 May 2015, we employed an ABAB dither pattern with $1\farcs25$ nod offset, and on 19 March 2016 we employed an ABBA dither pattern also with a $1\farcs25$ nod offset. The detection of the line in the two independent reductions demonstrates that its origin is not arising from systematic effects associated with dithering. Individual exposures were 180 seconds in duration during both nights, totaling approximately 2.30 hours on 27 May 2015 and 1.85 hours on 19 March 2016. The data from the two nights were reduced using the most updated version of the publicly available MOSFIRE data reduction pipeline (DRP) as of 19 March 2016. The reduction pipeline differences, stacks, and rectifies the nodded images, creating two-dimensional signal and noise spectra for each individual slit. Spectra from individual nights were reduced separately due to varying observing conditions and differences in the positions of the filler slits.

The average seeing was $0\farcs70$ on  27 May 2015 and $0\farcs85$ on 19 March 2016 in the J-band. Seeing on 27 May 2015 was calculated solely from the J-band alignment images directly preceding Y-band science observations. The seeing was stable on 27 May 2015 throughout our observations, but less stable on 19 March 2016. A bright star on the 19 March 2016 science mask allowed us to track the seeing in the Y-band during our observations, which we then normalized to the J-band seeing. Because the seeing differed between the two observing nights, we smoothed the 27 May 2015 two-dimensional spectra to match the seeing on 19 March 2016 before combining the spectra. The two individual two-dimensional spectra from the observations on 27 May 2015 and 19 March 2016 were combined by an inverse variance-weighted sum at each pixel. The inverse variance spectra were obtained from the two-dimensional noise spectra produced by the MOSFIRE DRP. 

The spectra from 27 May 2015, 19 March 2016, and the inverse variance stack of the two are shown in Supplementary Figure~\ref{fig:S1}. A single emission peak is observed in both the 27 May 2015 spectrum and the 19 March 2016 spectrum. The spatial and spectral location of the peak emission is slightly offset between the two observations despite using the same slit-mask setup. However, the offset is consistent with other emission lines in our slit-mask, once we degrade those emission lines to match the S/N of the {\lya} line. The two-dimensional stack shown in the bottom panel of Supplementary Figure~\ref{fig:S1} differs from the one shown in Figure 1. Supplementary Figure~\ref{fig:S1} shows a direct stack, whereas in Figure 1 we show a cross-correlated stack accounting for the observed vertical and horizontal shift. The cross-correlated spectrum is used for the calculation of the line flux and $S/N$. The observed wavelength of the {\lya} line is measured from the stack in Supplementary Figure~\ref{fig:S1}, which does not incorporate any spatial or spectral shifting.

To extract the one-dimensional spectrum from the two-dimensional co-added spectrum, we used an aperture whose size was chosen to enclose 2*FWHM of the average seeing during the observations. The one-dimensional spectrum shown in Figure 1 was extracted from the co-added two-dimensional spectrum, so the worse seeing was used to define the aperture size. At each point along the spectral axis, pixels within the spatial aperture were summed with weights determined from a gaussian whose width was also determined by the seeing, producing a one-dimensional spectrum. Flux calibration was done by comparing the one-dimensional extracted spectrum of a standard star we observed to a model spectrum of the same spectral type that was observed on a different night but with comparable atmospheric seeing and transparency. This step takes into account the slit-loss for a point source. Before comparison, we scaled the model spectrum to the apparent magnitude of the observed standard, then correct it for the airmass and galactic extinction of our observations. The ratio of the scaled, corrected model spectrum to the observed telluric spectrum represents the sensitivity function of the telescope. 

While the flux calibration corrects for slit-loss from a point-source, we expect that the {\lya} emission from \gname{} is more extended \cite{Steidel+11,Wisotzki+16}. The simplest model for the {\lya} spatial distribution is that it traces the rest-frame UV-continuum, sampled by the WFC3/IR filters. From the \HST{} F125W, F140W and F160W cutouts in Figure~3a, it is clear that the source is well-resolved. To estimate the flux lost from the slit due to the extended nature of \gname{} (in addition to the point-source term already accounted for), we compare the flux lost from the $0\farcs7$ MOSFIRE slit for our source versus a point source, both convolved with a Gaussian kernel whose width is determined by the seeing. We model our source with a Sersic profile using GALFIT \cite{Peng+10}, finding a best-fit Sersic index of $\sim2.7$ and effective (half-light) radius of $r_{\mathrm{eff}} \sim 1''$ in F140W, the \HST{} filter with the highest S/N. We find a slit-loss correction of $1.6$; i.e. we multiply our calibrated flux by this factor to account for the additional flux lost due to the extended nature of the source. This correction has been applied to all MOSFIRE flux and equivalent width values reported in the paper. 

Two bright galaxies can be seen near the top of the slit in Figure~3c. Their traces contaminate both spectra at approximately the same vertical position as the top negative residual from the emission line. We subtracted these traces and their negative residuals by calculating the average pixel value in each row of the reduced two-dimensional spectrum from each night separately. The spectra shown in the top two panels of Supplementary Figure~\ref{fig:S1} are post-subtraction. The co-added spectrum shown in the bottom panel of Supplementary Figure~\ref{fig:S1} was obtained from the trace-subtracted individual spectra. 

In principle, the continuum from this galaxy could be detected by binning over a sufficiently large bandpass. We note that the negative signals of the aforementioned contaminating galaxies are larger than the continuum predicted from the measured F125W HST flux, which is red-ward of {\lya} without containing any flux from the {\lya} line itself. We nonetheless checked for continuum from \gname{} by comparing the means of the one-dimensional spectrum red-ward and blue-ward of the {\lya} emission before subtraction of the contamination. When computing the means, we used approximately the same size bandpass red-ward and blue-ward of {\lya} ($\sim 550$\AA{}). Before computing the means we clip off $\sim10\%$ from both ends of the distributions to reject sky line contamination. We do not measure a statistically larger mean red-ward than blue-ward of {\lya}, which is evidence that we are not detecting the continuum from \gname{}. This is not surprising given the level of the contamination compared to the expected continuum signal. 

The $S/N$ of the emission line was determined from the one-dimensional spectrum shown in Figure 1. S is the flux of the line and N is the RMS noise summed in quadrature over the same bandpass used to determine the flux, which is $10501-10510$\AA{}. The line flux is \linefluxunits{}, resulting in a $S/N=6.7$ detection. The bandpass is limited on the red side due to the sky line as shown in Figure 1. The sky line renders an asymmetry test unreliable and also potentially reduces the line flux we are able to measure. The {\lya} detection is not sensitive to the pixel contributing a significant amount of flux just blueward of the sky line in Figure 1; the $S/N$ is $6.0$ if this pixel is not included in the bandpass. The emission line at $10504$\AA{} is the only $S/N > 5$ feature within the $\pm50$\AA{} window from GLASS that appears in both the 27 May 2015 and 19 Mar 2016 spectra.

\subsection*{Description of the GLASS data} 

GLASS is an \HST{} large program that targeted 10 massive galaxy clusters during Cycle 21 \cite{Schmidt+14,Treu+15a}. Data were taken with the WFC3 G102 and G141 grisms in the cluster cores and the ACS G800L grism in parallel fields. Because we focus on the cluster core in this work, we only use the data from the WFC3 grisms. 14 orbits were allocated for each cluster, distributed to achieve uniform sensitivity over the range in wavelengths $0.8-1.7\mu m$. Each cluster was observed at two position angles (P.A.s) approximately 90 degrees apart to aid in contamination subtraction. The two GLASS P.A.s for \MJ{} (P.A.$=8^{\circ}$ and P.A.$=88^{\circ}$) are shown in Figure~3c. 

The spectral extraction for \gname{} we performed is similar to that described by \cite{Schmidt+16}, but with two important differences that resulted in a slight enhancement in the S/N of the {\lya} emission line ($S/N = 2.4$ in this work versus $S/N = 2.1$ from \cite{Schmidt+16}). First, we used an updated reduction of the GLASS data. This is based on a more reliable detection and flux scaling of contaminating sources affecting the flux levels and noise properties of the contamination-subtracted spectra from which the line flux is estimated. Second, we force-extracted the spectrum at the exact coordinates of the source, and in doing so assumed a circular aperture for contamination modeling. In \cite{Schmidt+16}, we extracted the spectra using SExtractor-defined segmentation maps of the nearest matches to the coordinates of the object.

The G102 and G141 spectra at both P.A.s are shown in Figure 2. {\lya} is detected marginally at $S/N=2.4$ in the P.A.$=88^{\circ}$ G102 spectrum at $10500\pm50 $\AA{}, implying a redshift of $z=7.64\pm0.04$, in agreement with the redshift measured with Keck/MOSFIRE. The inferred {\lya} equivalent width from GLASS is $W_{\mathrm{Ly}\alpha} =27\pm11 $\AA{}. Because the continuum was not detected in the spectrum, the continuum flux density used for the equivalent width measurement was estimated from the broadband {\HST} flux in the F125W band. This band was chosen because it samples the rest-frame UV continuum near the {\lya} wavelength without containing any contribution from the {\lya} emission or the {\lya} break. 

The G141 spectral coverage allows us to search for additional emission lines at {\zlya}, such as {\civlambda} and {\ciiilambda}, both of which would be blended in the GLASS spectroscopy. No significant spectral features were recovered from this search. The $3\sigma$ limits on the fluxes and equivalent widths for {\civ} and {\ciii} are shown in Table 1. The continuum flux densities used to determine the equivalent widths were estimated in a similar manner to the {\lya} equivalent width measurement described above, but instead the F160W and F140W filter fluxes were used for {\civ} and {\ciii}, respectively. 

Nominally, the {\lya} flux derived from the grism is a factor of $\sim3$ larger than the Keck/MOSFIRE flux. However, the grism-derived flux we report in Table 1 was obtained using an aperture designed to optimize the signal to noise ratio ($S/N$) of the line, whereas the aperture of the MOSFIRE observations is fixed by the size of the slit, $0\farcs7$. Once the apertures are matched, the fluxes from the grism and MOSFIRE agree within the uncertainties. Higher levels of contamination are the most likely cause for the non-detection of {\lya} in P.A.$=8^{\circ}$ of the grism.

\subsection*{Alternative interpretations of the emission line} 

\begin{figure}
\centering
\includegraphics[width=0.99\textwidth]{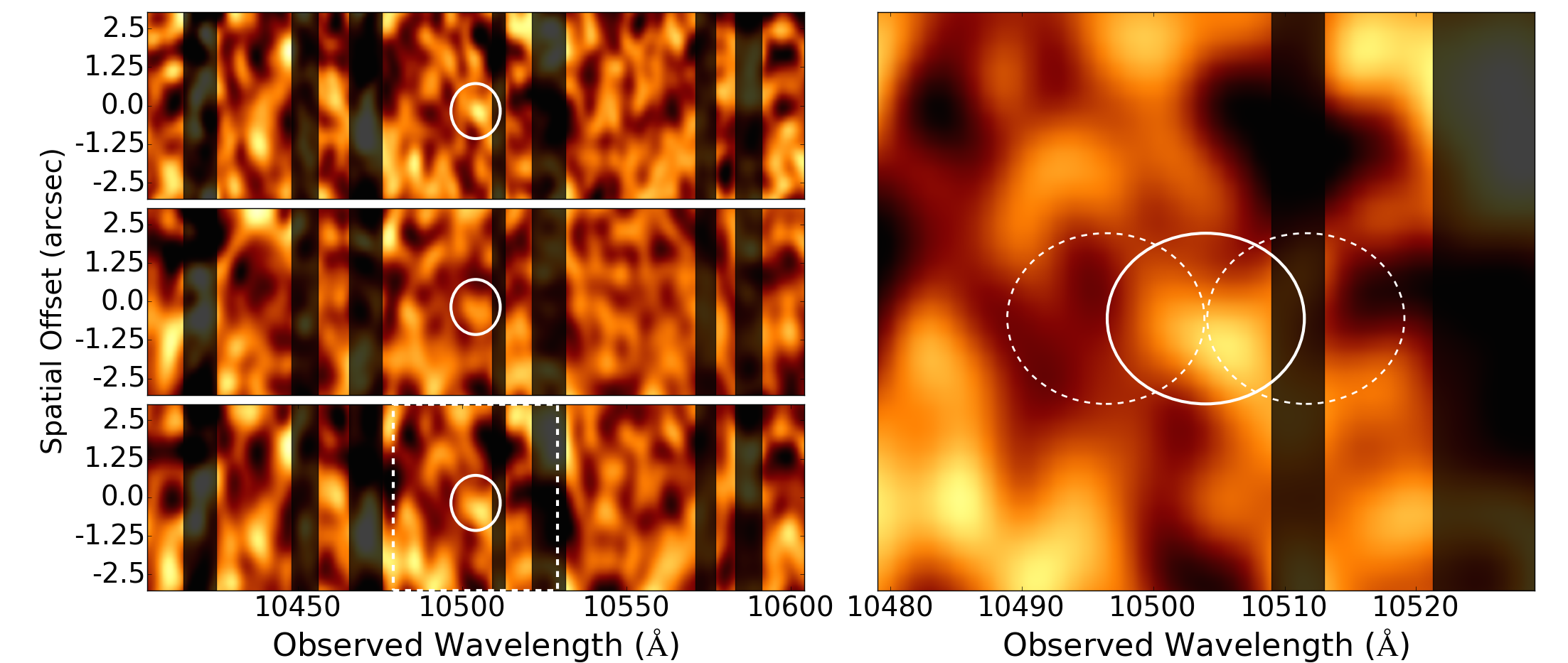}
\caption{Two-dimensional signal spectra from the (\textbf{Top left}) 2.30 hr 27 May 2015 observation, (\textbf{Middle left}) 1.85 hr 19 March 2016 observation and (\textbf{Bottom left}) direct inverse-variance stack of the two spectra. The emission line, which we are confident is the {\lya} line, is circled in white in all panels. (\textbf{Right}) Zoomed in region on the emission line from the stacked spectrum shown in the bottom left panel. Again, the emission line at $10504$\AA{} is circled in white. The dotted circle to the left (right) of the emission line shows the expected location of the $3726$\AA{} ($3729$\AA{}) line in the {\oii}$\lambda\lambda3726,3279$\AA{} doublet if the observed emission line was instead the $3729$\AA{} ($3726$\AA{}) line at $z=1.818$. The observed emission line at $10504$\AA{} cannot be the $3729$\AA{} line because we would have detected the bluer $3726$\AA{} line in that case. In the case that the observed emission line at $10504$\AA{} was the $3726$\AA{} line, the sky line would have obscured the $3729$\AA{} line. The grism spectra provide strong evidence against the {\oii} interpretation of the line, however. The spectra in all panels have been smoothed to the average seeing of each observation using a circular Gaussian kernel.  }
\label{fig:S1}
\end{figure}

We find that the most likely interpretation of the emission line is {\lya} at $z=7.640$. However, we explore the next most likely scenario in which the line is \oii{}$\lambda \lambda 3726,3729$ at $z=1.818$. The strong break observed in the photometry at $\sim1.05 \, \mu m$ rules out emission lines other than {\lya} and {\oii} since it is either the Lyman or Balmer break. The resolution of MOSFIRE is sufficient to resolve the \oii{} doublet at $z=1.818$. Supplementary Figure~\ref{fig:S1} shows the expected locations of the $3726$\AA{} and $3729$\AA{} lines in the {\oii} doublet if the emission line observed at $\sim10504$\AA{} was the $3729$\AA{} line or the $3726$\AA{} line, respectively. If the line at $\sim10504$\AA{} was the $3729$\AA{} line, then the bluer $3726$\AA{} line would have been observed since it falls in a region free of strong sky emission lines. The typical $3729$\AA{} $/ 3276$\AA{} intensity ratio for star-forming galaxies is $1.5/1$, so the $3726$\AA{} line would have been detected in this case. If the line at $\sim10504$\AA{} was the $3726$\AA{} line, then the redder $3729$\AA{} line would have fallen on a sky line and would therefore have been undetectable. 

The strongest evidence against the line being {\oii} comes from the grism observations. The GLASS grism spectra cover the \oiii{} pair in the case where the emission line at $\sim10504$\AA{} is \oii{}. However, the \oiii{} pair is not detected in either P.A. (Figure 2), resulting in a $2\sigma$ upper limit of {\othreetwo}$<0.65$ in P.A.=$88^{\circ}$. At $z=1.818$, the galaxy would have an intrinsic stellar mass of $\sim1\times10^{9} M_{\odot}$, which would imply {\othreetwo}$\geq1.5$ \cite{Henry+13}, inconsistent with our observations. Furthermore, the SED fitting strongly prefers the $z=7.640$ solution ($\chi^2_{\nu} = 0.9$) over the $z=1.818$ solution ($\chi^2_{\nu} = 3.4$). In summary, we are confident that the line at $\sim10504$\AA{} is not {\oii} and therefore must be {\lya}. 

\end{document}